\UseRawInputEncoding

\documentclass[%
        reprint,
        twocolumn,
        superscriptaddress,
        nofootinbib,
        aps
    ]{revtex4-2}
\usepackage{graphicx}

\usepackage{hyperref}
\hypersetup{
	colorlinks=true,
	linkcolor=blue,
	citecolor=blue,
	urlcolor=blue,
}

\usepackage{physics}
\usepackage{amsmath}
\usepackage{amssymb}
\usepackage{bm} 
\usepackage{bbm} 
\usepackage{mathtools}
\usepackage{dsfont}
\usepackage{braket}
\usepackage{tikz}
\usetikzlibrary{quantikz2}
\usepackage{url}

\usepackage{xspace}
\usepackage{orcidlink}

\renewcommand{\vec}[1]{\bm{#1}}

\renewcommand{\selectlanguage}[1]{}

\begin{document}
\title{Classification of the Fashion-MNIST Dataset on a Quantum Computer}

\author{Kevin Shen\,\orcidlink{0009-0005-2506-4056}}
\email{kevin.shen@bmwgroup.com}%
\affiliation{Technical University of Munich, TUM School of Natural Sciences, Physics Department, 85748 Garching, Germany}%
\affiliation{Munich Center for Quantum Science and Technology (MCQST), Schellingstr. 4, 80799 M{\"u}nchen, Germany}%
\affiliation{BMW Group Central Invention, 80788 M{\"u}nchen, Germany}%
\affiliation{applied Quantum algorithms (aQa), Leiden University, The Netherlands}

\author{Bernhard Jobst\,\orcidlink{0000-0001-7027-3918}}
\affiliation{Technical University of Munich, TUM School of Natural Sciences, Physics Department, 85748 Garching, Germany}%
\affiliation{Munich Center for Quantum Science and Technology (MCQST), Schellingstr. 4, 80799 M{\"u}nchen, Germany}%
\affiliation{BMW Group Central Invention, 80788 M{\"u}nchen, Germany}
\author{Elvira Shishenina}
\affiliation{BMW Group Central Invention, 80788 M{\"u}nchen, Germany}
\author{Frank Pollmann\,\orcidlink{0000-0003-0320-9304}\,}
\affiliation{Technical University of Munich, TUM School of Natural Sciences, Physics Department, 85748 Garching, Germany}%
\affiliation{Munich Center for Quantum Science and Technology (MCQST), Schellingstr. 4, 80799 M{\"u}nchen, Germany}

\date{\today}

\begin{abstract}
The potential impact of quantum machine learning algorithms on industrial applications remains an exciting open question. Conventional methods for encoding classical data into quantum computers are not only too costly for a potential quantum advantage in the algorithms but also severely limit the scale of feasible experiments on current hardware. Therefore, recent works, despite claiming the near-term suitability of their algorithms, do not provide experimental benchmarking on standard machine learning datasets. We attempt to solve the data encoding problem by improving a recently proposed variational algorithm \cite{dilip_data_2022} that approximately prepares the encoded data, using asymptotically shallow circuits that fit the native gate set and topology of currently available quantum computers. We apply the improved algorithm to encode the Fashion-MNIST dataset \cite{xiao_fashion-mnist_2017}, which we make openly available \cite{shen_classification_2024} for future empirical studies of quantum machine learning algorithms. We deploy simple quantum variational classifiers trained on the encoded dataset on a current quantum computer ibmq-kolkata \cite{ibmq} and achieve moderate accuracies, providing a proof of concept for the near-term usability of our data encoding method. 
\end{abstract}

\maketitle

\section{Introduction}
\begin{figure*}[t]
    \centering
    \includegraphics[width=0.9\linewidth]{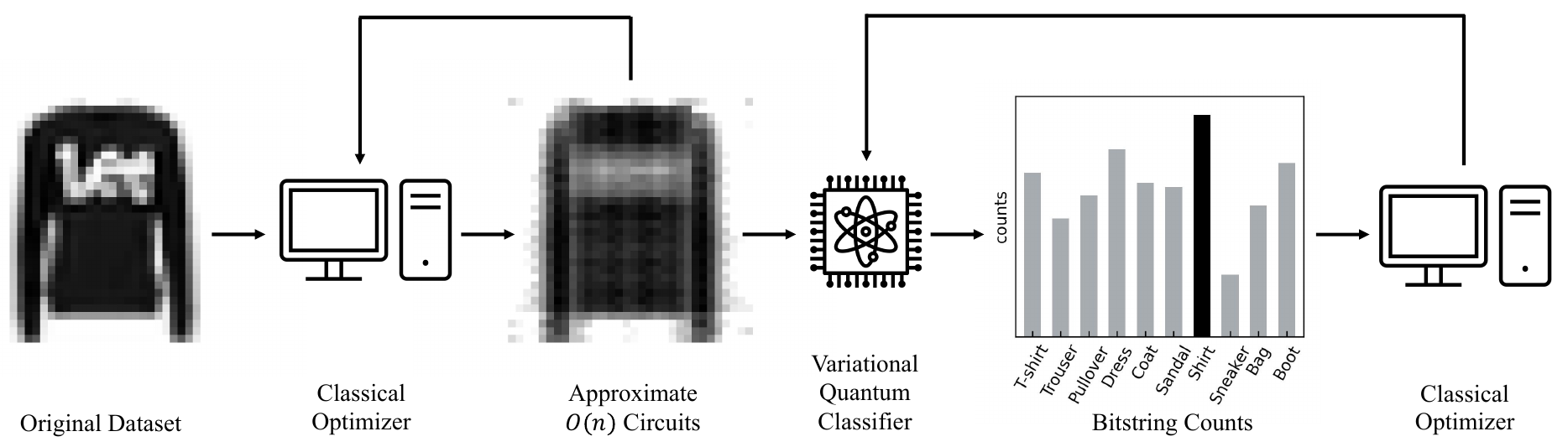}
    \caption{Schematic description of the contents of this work. On classical computers, we optimize the parameterized quantum circuits with gate complexities linear in the number of qubits that approximately prepare the Flexible Representation of Quantum Images (FRQI) of the Fashion-MNIST dataset. The encoded dataset is then used for training a variational quantum classifier on classical computers. We provide results from deploying the trained circuits on a quantum computer ibmq-kolkata.}
    \label{fig:schematic}
\end{figure*}

\par Supervised machine learning has been viewed as a potentially promising application for quantum computers \cite{dunjko_quantum-enhanced_2016, biamonte_quantum_2017, dunjko_machine_2018, carleo_machine_2019}. Many early supervised quantum machine learning (QML) algorithms \cite{, lloyd_quantum_2013, rebentrost_quantum_2014} that are based on the Harrow–Hassidim–Lloyd algorithm \cite{harrow_quantum_2009} and quantum random accessible memories \cite{giovannetti_quantum_2008} suggested potential speedups against the best-known classical algorithms. This in turn fostered the development of competing quantum-inspired classical algorithms \cite{tang_quantum_2021, gilyen_improved_2022}. Encouraged by the development of noisy intermediate-scale quantum (NISQ) \cite{preskill_quantum_2018} computers, supervised QML algorithms with parameterized quantum circuits (PQCs) \cite{peruzzo_variational_2014, mcclean_theory_2016, cerezo_variational_2021} including quantum kernels \cite{benedetti_parameterized_2019, havlicek_supervised_2019, schuld_quantum_2019} and variational classifiers \cite{benedetti_parameterized_2019, havlicek_supervised_2019, schuld_quantum_2019, lloyd_quantum_2020, jerbi_quantum_2023} have raised increasing attention. Such algorithms have been rigorously shown to solve certain classically intractable learning problems efficiently \cite{liu_rigorous_2021, huang_power_2021}. Nonetheless, whether supervised QML algorithms will also achieve a practical advantage in any industry-relevant application remains an exciting and open question, the answer to which presumably requires not only theoretical arguments but also empirical studies of benchmarking problems on real quantum computers.
\par As of today, the practical application of supervised QML algorithms faces several challenges, including the data encoding problem, i.e., the preparation of quantum states that represent the training and testing data, which is a prerequisite for any supervised learning task. Conventional methods for data encoding need to compromise between the number of qubits and circuit depth. At one extreme, product encoding \cite{stoudenmire_supervised_2017, schuld_quantum_2019} uses a tensor product circuit with just one single-qubit rotation per qubit but requires one qubit per dimension of data. At the other extreme, amplitude encoding \cite{lloyd_quantum_2013, schuld_quantum_2019, schuld_circuit-centric_2020} utilizes the superposition property of quantum states to encode the data in the amplitudes of a logarithmic number of qubits, yet demanding a circuit of depth exponential in the number of qubits \cite{vartiainen_efficient_2004, plesch_quantum-state_2011}. Recent works have also considered more general encoding methods \cite{schuld_quantum_2019}, best represented by the `data-reuploading' models \cite{ perez-salinas_data_2019, schuld_effect_2021, perez-salinas_one_2021}, in which the number of qubits and circuit depth can, in principle, be freely chosen. In these methods, the resource consumption should be nevertheless similar to the two cases above in order to ensure that different data points are mapped to different quantum states, which requires at least as many free parameters in the encoding circuits as the dimension of the data.
\par The data encoding problem not only acts as a bottleneck on the potential speedup in many supervised QML algorithms but also limits the scale of benchmarking problems that can be currently experimentally implemented.
So far most research either considered simple binary datasets \cite{grant_hierarchical_2018, havlicek_supervised_2019, schuld_quantum_2019,  schuld_circuit-centric_2020, bartkiewicz_experimental_2020, chalumuri_hybrid_2021, johri_nearest_2021, haug_quantum_2023, peters_machine_2021} or applied dimension reduction techniques such as principal component analysis or coarse-graining to pre-process the datasets \cite{grant_hierarchical_2018, schuld_circuit-centric_2020, kerenidis_quantum_2020, zhang_toward_2020, chalumuri_hybrid_2021, johri_nearest_2021, bokhan_multiclass_2022, hur_quantum_2022, lu_digital-analog_2024}. To the best of our knowledge, there has not been any experimental implementation of a supervised QML algorithm on a quantum computer considering a full-scale standard machine learning dataset. Recently, a variational algorithm inspired by the sequential generation of matrix product states (MPS) on quantum computers \cite{schon_sequential_2005, schon_sequential_2007, schollwoeck_density-matrix_2011, ran_encoding_2020, lin_real-_2021} that encodes image data in their Flexible Representation of Quantum Images (FRQIs) \cite{le_flexible_2011} with shallow optimized PQCs up to controllable approximation errors has been proposed \cite{dilip_data_2022}. Based on this algorithm, Ref. \cite{iaconis_tensor_2023} successfully demonstrated the approximate amplitude encoding of complex road image data on an ion-trap computer. This algorithm holds the broad potential to facilitate empirical studies of supervised QML algorithms on quantum computers in the near future. 
\par In this work, our contribution is threefold. First, we improve the resource efficiency of the PQCs in the above variational algorithm. By using only the typical native gates of current quantum computers in the ansatz, now image data are encoded to similar approximation errors at a smaller circuit depth, thus allowing subsequent data-processing tasks to be more faithfully implemented experimentally. Second, we apply the algorithm with the improved ansatz to encode the full Fashion-MNIST dataset \cite{xiao_fashion-mnist_2017} at different approximation accuracies. To facilitate future research, we make all circuits available in QASM \cite{cross_open_2017} file format on Zenodo \cite{shen_classification_2024}. Third, to showcase the near-term applicability of the encoded data, we train some simple variational classifiers on the encoded data to perform the standard ten-class classification task. The classifiers are first trained in simulation and then deployed on a state-of-art superconducting quantum computer, ibmq-kolkata \cite{ibmq}, via cloud access, where the best test accuracy of about $40\%$ is achieved, indicating that the images have been faithfully encoded despite noise. The data encoding PQCs and the classifiers are first trained on classical computers and then deployed on the quantum computer. Yet, this experiment serves as a proof-of-principle for empirical studies of supervised QML algorithms with full-scale standard machine learning datasets and acts as a stepping stone towards the full demonstration of supervised QML algorithms on more powerful quantum computers in the future. A schematic of the contents of this work is given in Figure \ref{fig:schematic}.
\par The organization of the paper is as follows. In Section \ref{sec:image representation}, we explain the encoding of the Fashion-MNIST dataset. In Section \ref{sec:classification}, we describe the multi-class classification experiment and present the results. We end with a discussion in Section \ref{sec:discussion}. 

\section{Efficient Encoding of the Fashion-MNIST Dataset}
\label{sec:image representation}
\subsection{Flexible Representation of Quantum Images}
\begin{figure*}[t]
\centering
\includegraphics[width=0.95\linewidth]{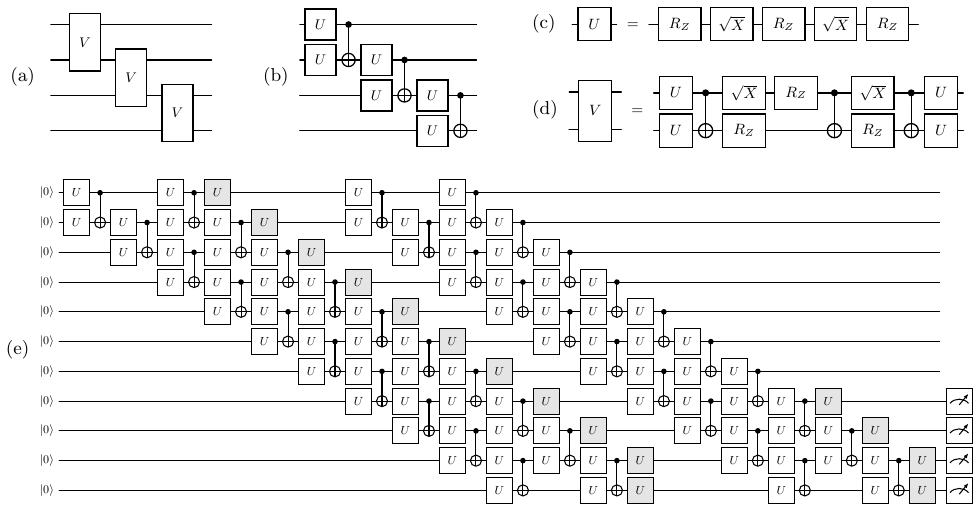}
\caption{(a) General ansatz with two-qubit gates $V$ arranged in a staircase pattern (same as in Ref. \cite{dilip_data_2022}). (b) Sparse ansatz with blocks of $2$ single-qubit gates $U$ and a CNOT gate arranged in a staircase pattern. (c) Decomposition of a $U$ gate on ibmq-kolkata. (d) Decomposition of a $V$ gate on ibmq-kolkata. (e) A circuit with $2$ layers of sparse ansatz for data encoding and $2$ layers of sparse ansatz for classification. Additional gates are colored in gray. The last $4$ qubits that are associated with the meter-like symbol are measured at the end of the circuit.}
\label{fig:ansätze}
\end{figure*}

Representing image data as quantum states is a prerequisite for image processing on quantum computers. Popular methods such as amplitude encoding \cite{lloyd_quantum_2013, schuld_quantum_2019, schuld_circuit-centric_2020}, the Novel Enhanced Quantum Representation of Digital Images (NEQR) \cite{zhang_neqr_2013} and the Flexible Representation of Quantum Images (FRQI) \cite{le_flexible_2011} use quantum states to represent image data of exponentially many pixels in a linear number of qubits by exploiting the superposition property of quantum states, making them particularly interesting for quantum computers of limited size that we have today. In this work, we choose to work with the FRQI for encoding the Fashion-MNIST dataset (below referred to as the dataset) for three reasons. First, the FRQI states of typical images are empirically found to possess lower entanglement entropy compared to states using one of the other two methods \cite{jobst_efficient_2023}, and thus can be approximated by shallower optimized PQCs to the same accuracy under the variational algorithm to be described in the next section. Second, we numerically find that the dataset is classified at higher accuracies when encoded as FRQI states compared to the other two methods (see Appendix \ref{app:supp_num_results}). Third, as the FRQI was also the choice of Ref. \cite{dilip_data_2022}, using it allows for a direct comparison.
\par Consider a grayscale square image with $2^n \times 2^n$ pixels, which we index as a $2^{2n}-$dimensional array $\vec{x} = (x_0, x_1, \dots, x_{2^{2n}-1})$, where $x_j \in [0,1]$ for $0 \leq j \leq 2^{2n}-1$ is the normalised color value of the $j$th pixel. Its FRQI is a quantum state $\ket{\psi}$ with $2n+1$ qubits described by
\begin{equation}
    \ket{\psi} = \frac{1}{2^n} \sum_{j=0}^{2^{2n}-1} \ket{j} \otimes \left( \cos(\frac{\pi}{2} x_j)\ket{0} + \sin(\frac{\pi}{2} x_j)\ket{1} \right).
    \label{eq:state_general_structure}
\end{equation}
The first $2n$ qubits capture the positions of the pixels in binary representation and the phase of the last qubit stores the color information. However, as shown in Ref. \cite{le_flexible_2011}, the exact preparation of an FRQI state requires a deep circuit with the number of CNOT gates scaling quadratically with the number of pixels, and hence exponentially in the number of qubits. Such a high gate complexity makes the exact encoding of the dataset infeasible in near-term experiments considering the noise levels of current quantum computers. 
\subsection{Approximate Preparation with Parametrized Quantum Circuits} 
\par We now introduce the variational algorithm for the approximate preparation of FRQI states. The main idea here is to encode each image with a separately optimized PQC. For each image, all parameters in the PQC are first randomly initialized and then iteratively updated by a classical optimizer to maximize $|\langle \tilde{\psi}| \psi \rangle |^2$, the overlap between the prepared state $\ket{\tilde{\psi}}$ and the target FRQI state $\ket{\psi}$. In Ref. \cite{dilip_data_2022}, the PQCs are constructed by repeatedly applying an ansatz (below referred to as the general ansatz) in which general parametrized two-qubit gates $V\in \mathrm{SU}(4)$ are arranged in a staircase pattern \cite{lin_real-_2021, barratt_parallel_2021, smith_crossing_2022}, as illustrated in Figure \ref{fig:ansätze}(a). The $V$ gate is defined by $15$ parameters as 
\begin{equation}
V(\vec{\theta}) = \text{exp}({-\frac{i}{2}\sum_{\alpha, \beta \in \{0,1,2,3\}^2 \backslash \{(0,0)\}} \hspace{-1cm} \theta_{\alpha, \beta} \sigma_\alpha \otimes \sigma_\beta}),
\label{eq:V}
\end{equation} where $\theta_{\alpha,\beta} \in \mathbb{R}$ are the tunable parameters, $\sigma_0$ is the identity matrix and $\sigma_1, \sigma_2, \sigma_3$ are the Pauli matrices. Previous numerical results show that only $2$ layers of the general ansatz are already sufficient to approximate the FRQI states of the dataset at above $95\%$ fidelity on average \cite{jobst_efficient_2023}. In addition, it has been observed that under such small approximation errors, the classification accuracies obtained by the subsequently trained classifiers are only slightly reduced compared to those trained on the exact FRQI states \cite{dilip_data_2022}.
\par The good approximation provided by such PQCs is partially related to the efficient MPS representations of image data \cite{jobst_efficient_2023}. Typical images including those in the dataset have fast-decaying Fourier coefficients, thus allowing their FRQI states to be truncated as matrix product states (MPS) of small bond dimensions under bounded errors. In principle, we could exactly prepare the MPS approximation for each image at some maximum bond dimension $\chi$ on a quantum computer by applying a sequence of multi-qubit gates with the supports on $\lceil \log_2 \chi \rceil + 1$ qubits \cite{schon_sequential_2005, schon_sequential_2007, lin_real-_2021, dilip_data_2022}. However, these circuits will have a large depth after compilation as each gate needs to be compiled into a long sequence of $\mathrm{poly}(\chi)$ native single- and two-qubit gates. 
\par Using PQCs with $l$ layers of the general ansatz should be viewed as a heuristic alternative that covers a resource-efficient subset of the MPSs with maximum bond dimension $2^l$ \cite{ran_encoding_2020, lin_real-_2021, dilip_data_2022}. The PQCs are resource-efficient in the sense that, asymptotically, the circuit depth grows only linearly in both the number of qubits $2n+1$ and the number of layers $l$. Furthermore, any linearly connected quantum computer can execute the PQCs with much smaller compilation overhead, as there are only two-qubit gates that act on adjacent qubits. However, these circuits are still too resource-demanding to be faithfully implemented on ibmq-kolkata, the superconducting quantum computer we consider. This is because the $V$ gates are not natively implementable, and must be decomposed into long sequences of $3$ native CNOT gates and $25$ native single-qubit gates \cite{vatan_optimal_2004, jurcevic_demonstration_2021}, as depicted in Figure \ref{fig:ansätze}(d).
\par Therefore, following the motivation to reduce gate complexity, we propose a sparse version of the PQCs. In the new ansatz (below referred to as the sparse ansatz), gate blocks consisting of two fully parametrized single-qubit gates $U\in \mathrm{SU}(2)$, each defined by three parameters as \begin{equation}
U(\Vec{\theta}) = \text{ exp}(-\frac{i}{2} \sum_{\alpha\in \{1,2,3\}} \hspace{-0.2cm} \theta_\alpha \sigma_\alpha)
\end{equation} and a CNOT gate are arranged in a staircase pattern, as shown in Figure \ref{fig:ansätze}(b). The sparse ansatz may also be repeated multiple times to achieve higher approximation capability. The $U$ gate will get decomposed into a sequence of at most $5$ native single-qubit gates when it is compiled on ibmq-kolkata, as shown in Figure \ref{fig:ansätze}(c). 
\par Since CNOT and the native two-qubit gates of many other current quantum computers such as CZ and XX belong to the category of perfect entanglers, i.e., those capable of mapping product states to maximally entangling states, and as all perfect entanglers are mutually convertible with additional single-qubit gates \cite{kraus_optimal_2001}, the post-compilation PQCs on these quantum computers will have similar gate complexities. Here, we estimate the gate complexity of a quantum circuit by counting the number of CNOT (equivalently native two-qubit) gates, since two-qubit gates dominate over single-qubit gates by about an order of magnitude in terms of execution time and error rate in most current quantum computers. In this context, with one-third of the circuit depth and two-fifths of the number of parameters, one layer of the sparse ansatz is much simpler than one layer of the general ansatz after compilation. In the next section, we numerically demonstrate that PQCs with the sparse ansatz are indeed more resource-efficient, i.e., able to achieve higher approximation accuracies with simpler post-compilation circuits, and thus are less prone to errors in experimental implementation. 

\subsection{Implementation}
\label{sec:comp imp}
\begin{figure}[!]
\includegraphics[width=\linewidth, trim={0 0 0 0}, clip]{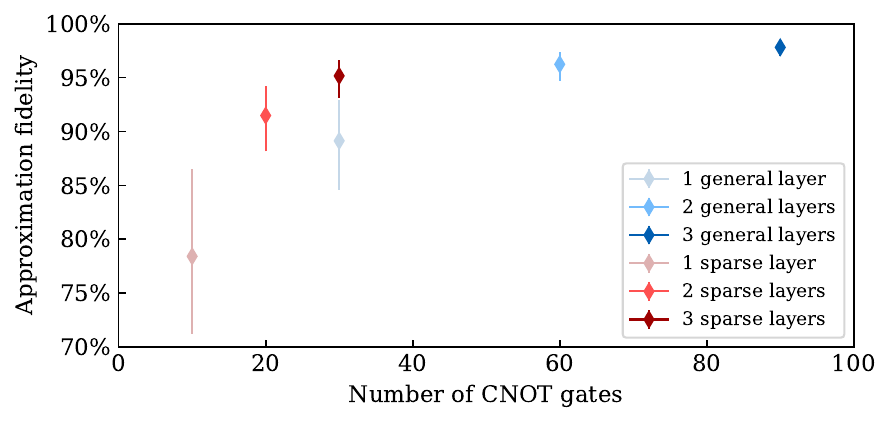}
    \caption{Fidelities between the target FRQI states and states approximately prepared by optimized PQCs of $1-3$ general or sparse layer(s). The markers and error bars show the mean values and the $25$th-$75$th percentile intervals computed over all $60000$ training images respectively.}
    \label{fig: compression results}
\end{figure}
We now apply the variational algorithm to encode the dataset. The dataset consists of $70,000$ labeled grayscale square images which have been randomly split into a training set $D^{\text{Train}}$ of size $M^{\text{Train}}=60,000$ and a test set $D^{\text{Test}}$ of size $M^{\text{Test}}=10,000$. Each set consists of an equal number of images belonging to $k=10$ different categories of fashion products. Each image, originally with $28 \times 28$ pixels, is first rescaled into $2^n \times 2^n = 32 \times 32$ pixels by bilinear interpolation and then flattened into a $1024-$dimensional array following a snake pattern as in Refs. \cite{dilip_data_2022, jobst_efficient_2023}. To make it explicit, let us write $D^{\text{Train}}=\{(\vec{x}^i, y^i) | 1 \leq i \leq M^{\text{Train}}\}$ ($D^{\text{Test}}$ likewise), where $\vec{x}^i \in [0,1]^{4^n}$ and $y^i \in \{0,\dots,k-1\}$ are the reshaped image arrays and their labels respectively.
\par To encode the dataset, we use PQCs acting on $(2n+1) = 11$ qubits as introduced in the previous section. We prepare three encoded datasets aiming for different approximation accuracies by using $1$--$3$ layers of the sparse ansatz respectively. In each case, we append an additional layer of single-qubit gates to the end of each PQC to enhance the approximation capability. An example two-layer PQC is given in Figure \ref{fig:ansätze}(e). For comparison, we also prepare three encoded datasets using $1$--$3$ layers of the general ansatz. The statistics of the approximation accuracies obtained in all six encoded datasets are visualized in Figure \ref{fig: compression results}. Results show that $3$ layers of the sparse ansatz give an average approximation accuracy of $95.1\%$, just $1\%$ lower than that of $2$ layers of the general ansatz. However, the former only requires half the number of CNOT gates and about one-third of the circuit depth compared to the latter and is thus more likely to be faithfully deployed on quantum computers. The details regarding the optimization process of the PQCs are discussed in Appendix \ref{implementation_details}.

\begin{figure*}[!]
\includegraphics[width=0.9\linewidth, trim={0 0 0 0}, clip]{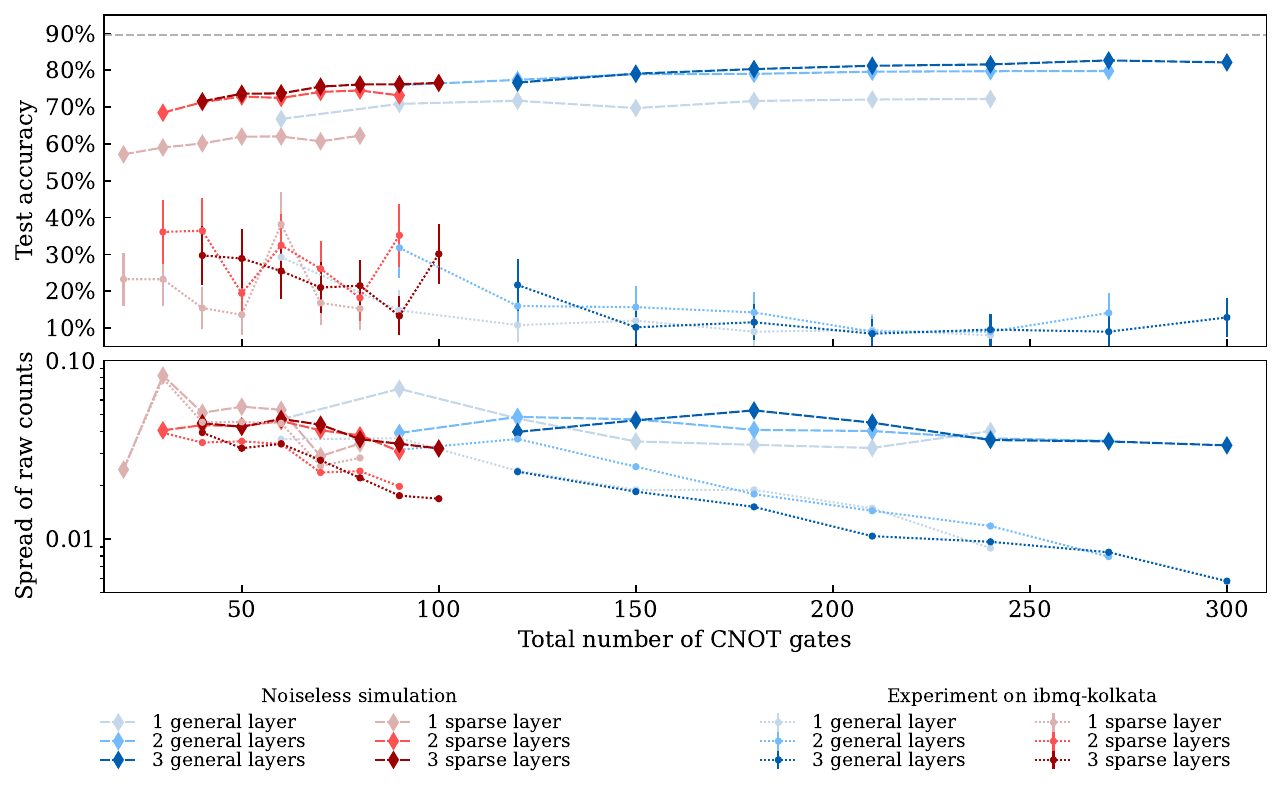}
    \caption{Test accuracies (top) and spreads of raw counts (bottom, log-scale) in noiseless simulations on classical computers (dashed curves) and in experiments (dotted curves) on the quantum computer ibmq-kolkata. Each curve corresponds to classifiers on the same encoded dataset, which are prepared by PQCs of $1$--$3$ sparse (left) or general (right) layer(s) respectively. On each encoded dataset, classifiers built from PQCs of $1$--$7$ layer(s) of the same ansatz as the data encoding PQCs are trained and tested. The horizontal axis labels the total number of CNOT gates, which estimates the circuits' gate complexities. Test accuracies in the simulation are computed over all $10000$ testing images. Test accuracies in experiments are computed over the first $100$ testing images. The error bars indicate the $95\%$ Wilson confidence intervals. Test accuracy from a support vector machine (see Appendix \ref{sec:Multiclass Classification Appendix}) of value $89.48 \%$ is displayed for reference (dashed gray line). The first $100$ testing images are used for computing the spread of raw counts in both simulations and experiments. The spread of raw counts, computed as the mean $L_2$ distance between each raw count vector and the center of the vectors, indicates the level of hardware noise experienced by the circuits.}
    \label{fig: simulation results}
\end{figure*}
\section{Multi-class Classification of the Fashion-MNIST Dataset}
\label{sec:classification}
\subsection{A Quantum Variational Classifier Trained in a Single Optimization}
\par We proceed to test the encoded datasets by solving the standard ten-class classification problem. In this task, the goal is to select a classifier $f$ from a hypothesis class that maximizes the test accuracy 
\begin{equation}\text{Acc}(f, D^{\text{Test}}) = \abs{\{(\vec{x},y) \in D^{\text{Test}} | f(\vec{x}) = y \} }/\abs{D^{\text{Test}}} \label{eq:acc}
\end{equation} which is an empirical estimation of the classifier's generalization performance on new images. 
\par There have been several QML strategies for tackling such multi-class classification tasks, which we discuss in Appendix \ref{sec:Multiclass Classification Appendix}. We attempt the problem by training a variational classifier that simultaneously distinguishes among images of all $k$ classes in a single optimization \cite{havlicek_supervised_2019, du_problem-dependent_2023, dilip_data_2022}. Such a classifier labels an image $\vec{x}$ by first transforming the encoded state $\ket{\tilde{\psi}}$ with a PQC $W(\vec{\theta})$ and then estimating the expectation values of $k$ different observables $O_i$ for $0 \leq i \leq k-1$, which we define to be in one-to-one correspondence with the $k$ classes, on the resultant state $W(\vec{\theta})\ket{\tilde{\psi}}$. The image $\vec{x}$ is assigned to the class corresponding to the maximum expectation value. The decision function can then be written as
\begin{equation}f(\vec{x}) = \arg \max_i \bra{\tilde{\psi}}W^\dagger(\vec{\theta})O_i W(\vec{\theta})\ket{\tilde{\psi}}. \label{eq: decision} \end{equation}
\par For demonstration purposes, we choose simple hypothesis classes consisting of PQCs with the same Ansätze as the data encoding PQCs so that the additional circuit depth is minimized. Just as for data encoding, we further append a final layer of single-qubit gates on the last $m=\lceil \log_2 k \rceil=4$ qubits, which are chosen to be the support of all $k$ observables $O_i$, to provide the flexibility of local basis transformation before measurements. An example circuit combining $2$ layers of sparse ansatz for data encoding and $2$ layers of sparse ansatz for classification is shown in Figure \ref{fig:ansätze}(e).
\par To save computational resources, we restrict ourselves to using mutually commutative observables $O_i$ of the following form \begin{equation}
O_i = \mathds{1} \otimes [\sum_{j=0}^{2^m} (A_{ij} + b_i)\ket{j}\bra{j}] \\ = \mathds{1} \otimes \sum_{j=0}^{2^m} A_{ij}\ket{j}\bra{j} + b_i \mathds{1} \label{eq:decision_obs} \end{equation} where $A$ is a $k \times 2^m$-dimensional weight matrix and $\vec{b}=(b_i)_{1\leq i \leq k}$ is $k$-dimensional bias vector. We remark that if we fix the parameters as $A_{ij}=\delta_{ij}$ and $\vec{b}=0$, which would enforce the constraints that the observables $O_i$ are orthogonal, we would recover the setup of Ref. \cite{dilip_data_2022} (see Appendix \ref{app:supp_num_results}). Instead, we allow both $A$ and $\vec{b}$ to be fully parametrized and optimizable. As all observables $O_i$ are diagonalized in the computational basis, they can be simultaneously evaluated on one quantum state. Hence, all parameters in $A$ and $\vec{b}$ can be updated together with the gate parameters $\vec{\theta}$ in a single optimization loop.
\subsection{Implementation}
\label{sec: class imp}
\par In this setting, we simulate the training of variational classifiers on classical computers using the training sets of the six encoded datasets respectively. On each encoded dataset, we train classifiers built from PQCs with $1$--$7$ layers of the same ansatz as the data encoding PQCs, amounting to $42$ classifiers in total. We then apply each trained classifier to classify all the images in the corresponding test set $D^{\text{Test}}$ to compute the test accuracy $\text{Acc}(f,D^{\text{Test}})$ (Equation \ref{eq:acc}). We observe that the test accuracy generally improves either when the data are encoded at higher approximation fidelities by PQCs with more layers of ansatz, or when the encoded data are classified by PQCs with more layers of the ansatz. We also show the test accuracy of a support vector machine trained on the exact FRQI states (see Appendix \ref{implementation_details}), which serves as an estimate of the performance of a well-designed classifier. Then we deploy the data encoding circuits together with the trained classification circuits on ibmq-kolkata. The details regarding the simulation and experiment processes are provided in Appendix \ref{implementation_details}. All simulation and experiment results are plotted in Figure \ref{fig: simulation results}. 
\par Due to the limited access to ibmq-kolkata, we only compute the test accuracies based on the first $100$ images. Despite the noise present in ibmq-kolkata, we still achieve
the best test accuracy of about $40\%$. We observe generally lower test accuracies from circuits with more CNOT gates in total, suggesting the benefits of the sparse ansatz. However, the trend is not clearly monotonic since multiple factors resulting from the complicated training process influence the experimental test accuracies. For example, an image that is correctly classified in simulation is more robust against hardware noise and sampling error (and hence more likely to be classified in experiments) if it has a larger margin, i.e., by how much the post-processed count corresponding to the correct class exceeds the others. 
\par To see that the higher number of CNOT gates, and hence larger hardware noise is a primary cause of the decreasing test accuracies in experiments, we need to look at an alternative metric that is not trivially related to test accuracy. We propose to compute the spread of raw counts, i.e., the average $L_2$ distance between the raw counts of a test image and the average raw counts over all $100$ test images, \begin{equation}
\frac{1}{100} \sum_{j=1}^{100} \norm{\vec{s}^j - \frac{1}{100} \sum_{i=1}^{100} \vec{s}^i}_2, \end{equation} which can be understood as a high dimensional analog of the standard deviation. By intuition, we know that a cluster of pure states, regardless of how they are distributed, always has a higher spread of raw counts than it would have if there were additional hardware noise --- for simplicity, consider a global depolarizing channel. No clear pattern is observed in simulations as there is no hardware noise. In contrast, in experiments, we observe that with more CNOT gates in total, the hardware noise increases. Then, the quantum states after the classification circuits experience heavier decoherence and get closer to each other, resulting in smaller spreads of raw counts.

\section{Discussion}
\label{sec:discussion}
In this work, we improve the resource efficiency of a previously proposed approximate data encoding algorithm \cite{dilip_data_2022} suitable for NISQ computers. With the improved algorithm, we fully encode the Fashion-MNIST dataset at different approximation accuracies, which can be used for future empirical studies of quantum image processing algorithms. Furthermore, from experiments on the quantum computer ibmq-kolkata with simple variational classifiers, we observe evidence that the data have been faithfully encoded and classified which is otherwise challenging without our improvements.
\par We note that in the long run, as error correction techniques mature, proneness to errors due to the high gate complexity of data encoding circuits would no longer be such a high-priority concern. Nonetheless, the approximated encoding of the Fashion-MNIST dataset we provide should remain valuable as it significantly saves the usage of quantum resources, which are likely to remain scarce and expensive, every time the dataset is used for benchmarking a quantum algorithm. We do not focus on the search for a variational classifier with potential quantum advantage, which has been the focus of many recent works and is left as a topic complementary to the data encoding problem.
\par Lastly, we would like to point out extensions and improvements to the data encoding algorithm, which may be case-dependent on the target quantum computer and the dataset that are worth investigating in the future. For instance, for data of much larger dimensions than that of the Fashion-MNIST dataset, one could try to further reduce the gate complexity of the data encoding PQCs by considering log-depth ansätze mimicking the preparation of normal MPS \cite{malz_preparation_2024}. Also, one may consider differently-shaped ansätze with similar gate complexity but potentially higher approximation capabilities if one uses quantum computers with circular or all-to-all connectivity. A remaining question is the search for a practical algorithm for optimizing the PQCs on quantum computers to encode data that are too large to be simulated.

\section{Ackownledgements}
The authors thank Yu-Jie Liu for insightful discussions and Carlos A. Riofr\'{i}o for collaboration on a related project. F.P. thanks Rohit Dilip, Yu-Jie Liu and Adam Smith for collaboration on a related previous project.
This research was funded by the BMW Group. E.S. is partly funded by the German Ministry for Education and Research (BMB+F) in the Project QAI2-Q-KIS under Grant 13N15583. F.P. acknowledges the support of the Deutsche Forschungsgemeinschaft (DFG, German Research Foundation) under Germany’s Excellence Strategy EXC-2111-390814868, the European Research Council (ERC) under the European Union’s Horizon 2020 research and innovation program (grant agreement No. 771537), as well as the Munich Quantum Valley, which is supported by the Bavarian state government with funds from the Hightech Agenda Bayern Plus.
K.S. thanks the Quantinuum Nexus team for providing access to ibmq-kolkata. 
\par \textbf{Data availability:} The .qasm files for quantum circuits approximately encoding the Fashion-MNIST are publicly available on Zenodo \cite{shen_classification_2024}.
\label{sec:ack}

\bibliography{ref, references}

\clearpage

\appendix
\section{Extended Discussion on Multi-class Classification}
\label{sec:Multiclass Classification Appendix}
\par As discussed in Refs. \cite{havlicek_supervised_2019, schuld_quantum_2019}, PQCs can be used to construct two different types of linear classifiers, i.e., models that separate data with hyperplanes. One of them, called the quantum kernel, solves the classification problem in its dual form --- it uses a quantum computer to evaluate the overlaps between data-encoded quantum states to estimate a kernel and then solves the remaining quadratic programming problem classically. In contrast, the other model, called the quantum variational classifier, attempts the problem in its primal form --- it directly parametrizes a hyperplane constructed from a PQC followed by measurements. For a fixed data-encoding method, the solution of a quantum kernel is guaranteed to be optimal in terms of training accuracy as a consequence of the representer theorem \cite{schuld_quantum_2019}, which however may be outperformed by variational classifiers in terms of generalization performance \cite{jerbi_quantum_2023}. 
\par The two algorithms above are tightly connected to their classical counterpart, the support vector machine (SVM) \cite{vapnik_nature_2000, burges_tutorial_1998}. A more formal mathematical treatment of the two quantum algorithms in the framework of SVMs is explored in Refs. \cite{havlicek_supervised_2019, liu_rigorous_2021}, but we will discuss the SVM from a practical perspective. An SVM solves the following minimization problem \begin{equation} 
\label{eq:svm1}
\min_{w,b,\xi} \norm{\vec{w}}^2_2 + C \sum_i \xi_i,
\end{equation}
subject to $\xi_i \geq 1-y_i(\langle \vec{w} , \vec{x_i} \rangle +b)$ and $\xi_i \geq 0$,
where $\vec{w}$ is the normal of the hyperplane, $b$ is the bias, $(\vec{x_i}, y_i)$ are the labeled raw data or data transformed by a feature map such as a quantum encoding and $\xi_i$ are slack variables for treating linearly nonseparable data. The first term $\norm{\vec{w}}_2^2$ inversely relates to the margin, so an SVM attempts to construct a hyperplane that maximizes the margin. The second term $C \sum_i \xi_i$ acts as a penalty term for mislabling. The problem may be reformulated as
\begin{equation}
\label{eq:svm2}
\min_{w,b,\xi} \frac{1}{C} \norm{\vec{w}}^2_2 + \sum_i \max(0, 1-y_i(\langle \vec{w} , \vec{x_i} \rangle +b)),
\end{equation}
where now the interpretation would be that the first term serves as a $L_2$ regularization and the second term as the total hinge loss, demonstrating the trade-off between margin maximization and empirical loss minimization.
\par An SVM is by nature an algorithm for binary classification tasks. Heuristic proposals that extend SVMs to multi-class classification tasks by leveraging a collection of SVMs \cite{hsu_comparison_2002} can also be directly applied to quantum kernels and quantum variational classifiers. Among them, the one-vs-one strategy consists of $k(k-1)$ SVMs running over all pairs of classes, and assigns an image to the class with the maximum `vote'. The one-vs-all strategy consists of $k$ SVMs, each distinguishing one class from the other $k-1$ classes combined, and assigns an image to the class with the maximum margin. These algorithms are heuristic in the sense that the SVMs are trained individually before being combined and it is generally not guaranteed that they are optimal for the holistic $k$-class classification task.
\par Alternatively, Ref. \cite{weston_support_1999} proposes an SVM-type multi-class classifier that can be trained in a single optimization with the decision function identical to that of the one-vs-all strategy:
\begin{equation}
    f(x) = \arg \max_l \langle \vec{w_l} , \vec{\vec{x}}\rangle+b_l.
    \label{eq:ovr}
\end{equation} Classifiers of this type are also popular in the QML literature. The decision functions of these classifiers
\begin{equation}
\begin{aligned}
    f(\vec{x}) &= \arg \max_i \bra{\tilde{\psi}}W^\dagger(\vec{\theta})O_i W(\vec{\theta})\ket{\tilde{\psi}}  \\
    &= \arg \max_i \Tr[W^\dagger(\vec{\theta})O_i W(\vec{\theta}) \ket{\tilde{\psi}}\bra{\tilde{\psi}} ]
\end{aligned}
\end{equation}
take exactly the same form as Equation \ref{eq:ovr} after identifying $\Tr[\cdot,\cdot]$ as the Hilbert-Schmidt inner product on the $4^n$-dimensional Hilbert space of Hermitian matrices, $\ket{\tilde{\psi}}\bra{\tilde{\psi}}$ as the transformed data, the weight of the identity component of $W^\dagger(\vec{\theta})O_i W(\vec{\theta})$ as the bias, and the orthogonal components of it as the normal vector. Different choices of $O_i$ can be found in the literature. Ref. \cite{havlicek_supervised_2019} considers both Pauli-Z observables and commuting Pauli observables, e.g., stabilizers. Ref. \cite{du_problem-dependent_2023} considers more general Pauli-based measurement observables. Ref. \cite{dilip_data_2022} considers projectors onto computational-basis states as observables. In this work, we consider its extension to trainable linear combinations of projectors as observables as discussed in the main text.
\par It is tempting to say that the total usage of quantum computers of these single-optimization classifiers is smaller than the heuristic approaches since the number of optimization loops does not scale with the number of classes $k$. However, this is not necessarily true, as the total usage of computers is a combined result of the circuits' gate complexities and the training process. Specific to quantum classifiers, due to the probabilistic nature of quantum measurements, it is imperative to perform enough repeated measurements (shots) such that the sampling error subsides the margins among classes. This problem is partially addressed in the ``Multiple Decision Problem" from statistics since the $1950$s \cite{bechhofer_single-sample_1959}. It has been found that a linear scaling $\mathcal{O}(k)$ in measurements is required for a fixed success probability against sampling error. Therefore, in summary, choosing the optimal strategy for multi-class classification with SVM-type quantum classifiers is generally not a trivial task, and should be problem-dependent. 
\section{Implementation Details}
\label{implementation_details} 
\begin{figure*}[!]
\includegraphics[width=0.8\linewidth, trim={0 0 0 0}, clip]{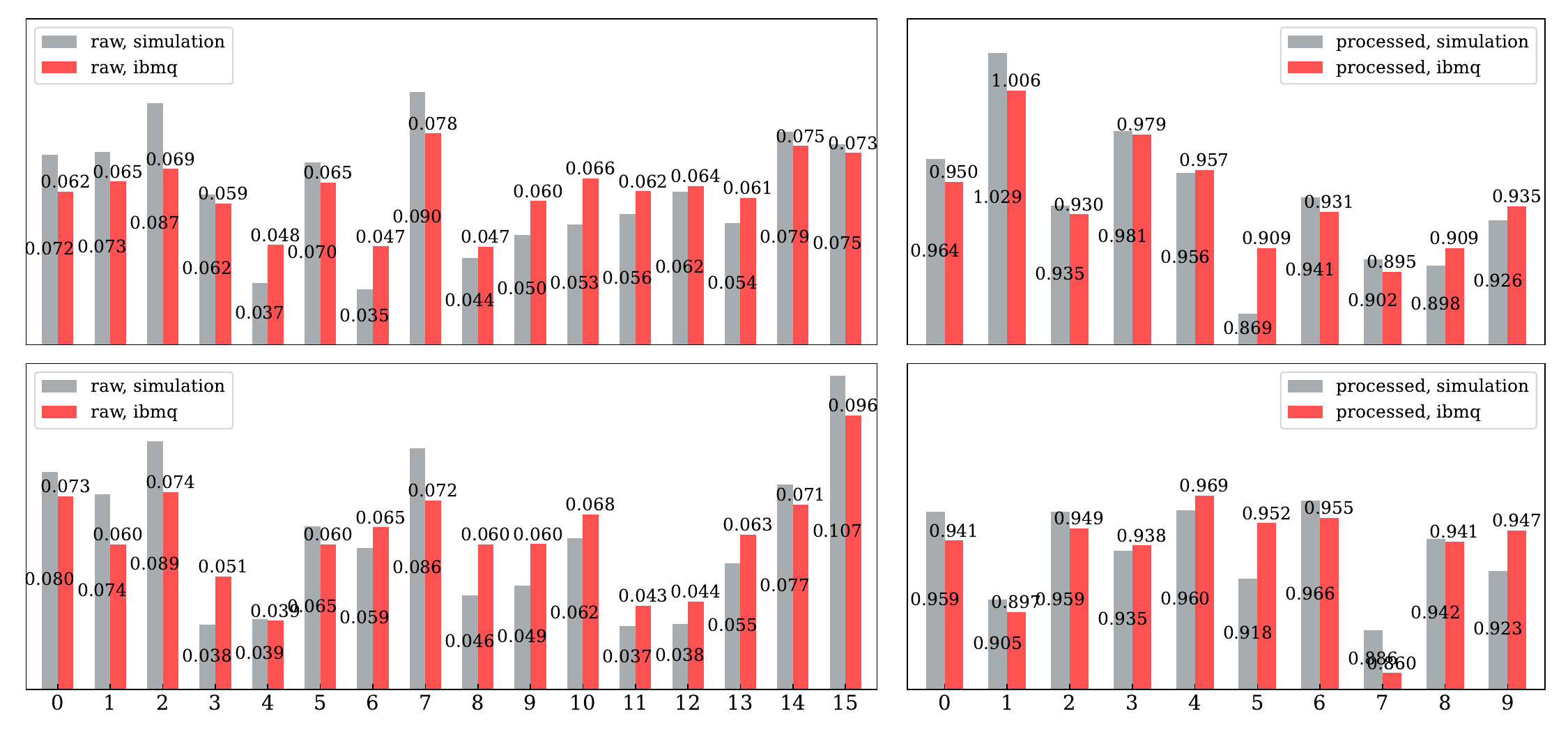}
    \label{fig: raw_bad}
\caption{Example measurement statistics. Left: Raw Counts. Right: Post-processed counts. Top: A successful example where the correct label `1' is found in both simulation and experiment. Bottom: A failed example where the correct label `6' is found in simulation, but the false label `4' is given in experiment primarily due to hardware noise.}
\end{figure*}
Here we discuss the implementation details for data encoding in Section \ref{sec:comp imp}. For each image, a PQC is first applied to the initial state $\ket{0}^{\otimes N}$ with all parameters uniformly randomly initialized in $[0,1]$. Then the parameters are simultaneously updated within the open real intervals by the Adam optimizer \cite{kingma_adam_2017} at a learning rate of $0.03$ for $10,000$ epochs, according to the loss function defined as $1- |\braket{\tilde{\psi}| \psi}|^2$, the infidelity between the prepared state and the target FRQI state. We would like to point out that we have also tried an adapted optimization algorithm initially designed for tensor networks (Algorithm $1$ in Ref. \cite{lin_real-_2021}). While this algorithm achieves competitive performance against the Adam optimizer when optimizing PQCs with the general ansatz, it is more prone to local minima for PQCs with more than $3$ layers of the sparse ansatz. We use TensorFlow \cite{tensorflow2015-whitepaper} for loading and rescaling the image data, Pennylane \cite{bergholm_pennylane_2018} for simulating quantum circuits, and Jax \cite{jax2018github} for optimization.
\par Now we discuss the implementation details for classification in Section \ref{sec: class imp}. To train a classifier on a collection of encoded states, all parameters $(\vec{\theta}, A, \vec{b})$ are uniformly randomly initialized in $[0,1]$. Then, the parameters are simultaneously updated within the open real intervals by the Adam optimizer at a learning rate of $8\times 10^{-4}$ for $200$ epochs over the whole training set $D^{\text{train}}$, under a log-softmax loss function defined below:
\begin{equation}
    L(f,D^{\text{train}}, B) = -\sum_{(\vec{x}, y) \in B}\log \frac{e^{C \tilde{\vec{s}}_y}}{\sum_{i=0}^{k-1} e^{C\tilde{\vec{s}}_i}}.
    \label{eq:cost_function}
\end{equation}
Here $C=128$ is a rescaling constant, $\tilde{\vec{s}}_y$ is the count of the correct label $y$ of the image $\vec{x}$ and $B \subset D^{\text{Train}}$ is a mini-batch of size $\abs{B}=80$. At each epoch, $D^{\text{Train}}$ is randomly divided into $60,000/80= 750$ mini-batches, i.e., $750$ subsets of $80$ images. The parameters are updated for $750$ times by evaluating the loss function once on each mini-batch to complete one epoch. We again used Pennylane and Jax for simulation and optimization. 
\par For experiments on ibmq-kolkata, we use the Qiskit Runtime Sampler primitive service \cite{ibmq} to execute quantum circuits. We have attempted different configurations and finally decided to use the default setting (optimization level $=3$ and the resilience level $=1$). Unlike in simulations, as we do not have access to the exact expectation values, we cannot exactly evaluate the decision functions (Equation \ref{eq: decision}). Instead, we need to process the measurement statistics. We use the maximum possible of $100,000$ shots to minimize the sampling error, and to better observe the effects of hardware noise. 
\par More precisely speaking, to classify an image, we prepare $N$ copies of the state $W(\vec{\theta})\ket{\tilde{\psi}}$, measure the last $m$ qubits of each copy in the computational basis, count the occurrences of the $2^m-1$ different bit strings that we write as $\vec{s} = (s_0, \dots, s_{2^m-1})$ with $\sum_{i=0}^{2^m-1}s_i=N$, and then post-process the counts as $\tilde{\vec{s}} =A\vec{s}+\vec{b}$. The image is then assigned to the class corresponding to the maximum post-processed value, $\arg \max_i \tilde{\vec{s}}_i$, which approximates the exact decision function up to sampling error. To better visualize the process, we show in Figure \ref{fig: raw_bad} the raw and post-processed counts for an image that is classified correctly both in simulation and in experiment, and for an image that is classified correctly in simulation but wrongly in experiment. Due to restricted access to ibmq-kolkata, the experiment is discontinuously conducted in several sessions during which the performance of ibmq-kolkata fluctuates due to regular recalibration.
\par Lastly we discuss the implementation of the SVM classifier. We use a kernel defined as $\braket{\psi(\cdot)|\psi(\cdot)}^2$ for the SVM classifier. Notice that we drop the $|\cdot|$ sign which is otherwise present in the definition of the fidelity, since the FRQI states $\ket{\psi}$ are real vectors by definition. As discussed in Appendix \ref{sec:Multiclass Classification Appendix}, such an SVM classifier gives an upper bound on the training accuracy for any variational classifier when the images are exactly encoded as FRQI states. To evaluate such an SVM classifier, we first prepare the exact FRQI states in vector form and apply the quadratic SVM from scikit-learn \cite{scikit-learn}, by selecting the polynomial kernel $(\gamma \langle \cdot, \cdot \rangle + r)^d$, choosing the parameters $\gamma=1, r=0, d=2$ and using the one-vs-rest strategy. After a grid search on the regularization constant (Equation \ref{eq:svm1} and Equation \ref{eq:svm2}), we find the best performance when setting $C=1000$ and plot it in Figure \ref{fig: simulation results}. 

\section{Supplemental numerical results}
\label{app:supp_num_results}
In the main text, we mentioned that one of the reasons we choose to work with FRQI states is that we empirically find them to be more accurately classified compared to other data encoding methods such as amplitude encoding or the NEQR. Here we present some additional numerical results. We first compare the classification accuracies with different data encoding methods. Then we study the effect of adding the post-processing method used in the main text by comparing it to the case of simply using the raw counts in the decision rules. Lastly, we compare the test accuracies achieved by variational classifiers using different ansätze. Unlike in the main text, all results presented in this section are based on the exactly encoded dataset, i.e., not approximately prepared by PQCs, and are obtained purely from classical simulations.
\par First, we will compare different ways of encoding the image data. We consider 3 different encodings, namely amplitude encoding~\cite{lloyd_quantum_2013, schuld_quantum_2019, schuld_circuit-centric_2020}, the FRQI~\cite{le_flexible_2011} and the NEQR~\cite{zhang_neqr_2013}. They all share the property that they encode the pixel values of an exponentially large image in a linear number of qubits. Consider a $2^n\times2^n$-pixel grayscale image (in our case $2^n=32$ as discussed in the main text) with the $j$th pixel value given by $x_j\in[0,1]$. Amplitude encoding is the simplest form of mapping the data to a quantum state, by simply choosing the amplitudes of a $2n$-qubit state proportional to the pixel values and then normalizing the state, i.e.,
\begin{equation}
    \ket{\psi} = \frac{1}{\sqrt{\sum_{j=0}^{2^{2n}-1} x_j^2}} \sum_{j=0}^{2^{2n}-1} x_j \ket{j}. 
\end{equation}
The FRQI works similarly, but instead uses an extra qubit to store the color information, with the state given by
\begin{equation}
    \ket{\psi} = \frac{1}{2^n} \sum_{j=0}^{2^{2n}-1} \ket{j} \otimes \left(\cos(\frac{\pi}{2}x_j) \ket{0} + \sin(\frac{\pi}{2}x_j) \ket{1}\right).
\end{equation}
Finally, the NEQR assumes that the pixel values are given by a bitstring of $q$ classical bits (i.e., $x_j = x_{j,0}x_{j,1}\ldots x_{j,q-1}$ with $x_{j,\ell}\in\{0,1\}$) and uses $q$ qubits to store the color information. The full NEQR state on $2n+q$ qubits is given by
\begin{equation}
    \ket{\psi} = \frac{1}{2^n} \sum_{j=0}^{2^{2n}-1} \ket{j} \otimes \ket{x_{j,0}x_{j,1}\ldots x_{j,q-1}}.
\end{equation}
Here, we will restrict the number of color qubits between one and three, such that the number of qubits remains comparable to the other encodings. Since the pixel values of image data are usually given with $8$-bit accuracy, we first coarse-grain the color resolution to the appropriate number of color qubits.

\begin{figure}[t]
    \centering
    \includegraphics{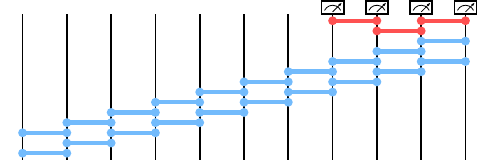}
    \caption{A one-dimensional sequential circuit as a variational classifier. The dumbbell-shaped symbols represent general parametrized two-qubit gates $V \in U(4)$, which are optimized during training. Unlike in the main text, here we also include a readout layer of gates that are colored in red, as it was originally introduced in Ref.~\cite{dilip_data_2022}. The last four qubits that are marked with the meter-like symbol are measured at the end of the circuit.}
    \label{fig:1d_sequential_classifier}
\end{figure}
\par Here we use the same setup for classification as originally proposed in Ref.~\cite{dilip_data_2022}. The circuits are slightly different from the PQCs using the general ansatz in the main text, with an additional readout layer of general parameterized two-qubit gates $V \in U(4)$
acting on the last four qubits that are to be measured, as shown in Figure~\ref{fig:1d_sequential_classifier}. We identify the first $10$ outcome bitstrings of the four measured qubits to be in one-to-one correspondence with the $10$ classes, and assign an image to the class corresponding to the maximum counts. The decision rules are the same as in the main text (Equation \ref{eq: decision}, \ref{eq:decision_obs}), except that we do not post-process the counts. (Equivalently, we fix the post-processing units as $A_{ij}=\delta_{ij}$ and $\vec{b}=0$).
\par For training the classifier, we initialize all gates as random unitaries close to the identity. The unitary gates are optimized using the Riemannian Adam algorithm implemented by the QGOpt library~\cite{qgopt, Luchnikov2021, Oza2009} as an extension of TensorFlow~\cite{tensorflow2015-whitepaper}. We use the same log-softmax loss function as in Equation~\ref{eq:cost_function}, except that we use the raw counts $\vec{s}$ instead of post-processed counts $\tilde{\vec{s}}$, and we treat the constant $C$ as another variational parameter. We start with a learning rate of $10^{-3}$ and train for $500$ epochs, then perturb the trained parameters to potentially get out of local minima and decrease the learning rate, before training for another $500$ epochs. We repeat this procedure until the test accuracy converges at a learning rate of $10^{-4}$. We use this procedure for training all classifier circuits in this section.
\begin{figure}[t]
    \centering
    \includegraphics[width=\linewidth]{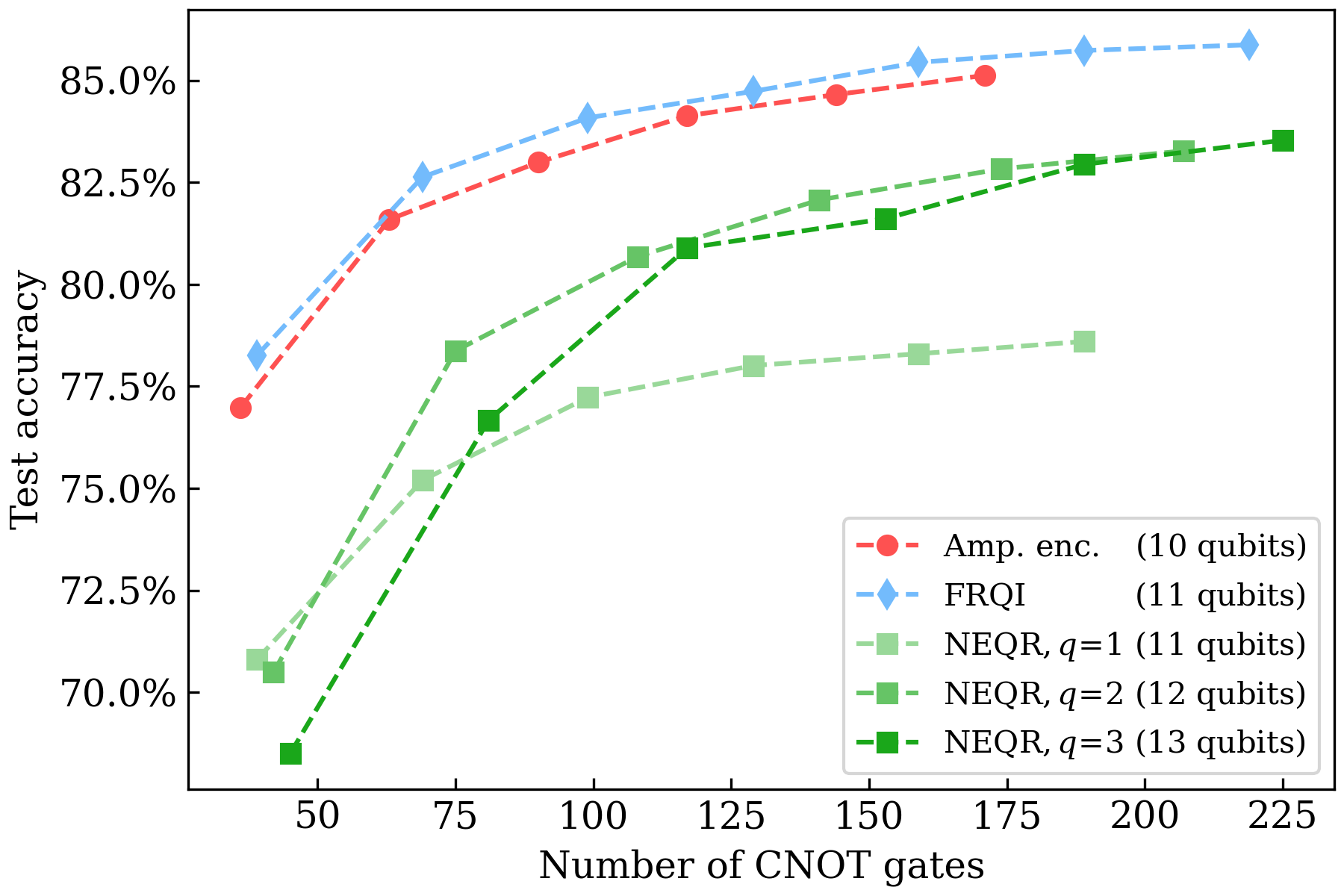}
    \caption{Comparing the test accuracy that one-dimensional sequential circuits achieve as variational classifiers on the exactly encoded dataset using different encoding methods. We consider amplitude encoding, the FRQI, and the NEQR with $1$--$3$ color qubits. Note that the different encodings may use a different number of qubits. We plot the test set accuracy against the number of CNOT gates in the classifier circuit.}
    \label{fig:test_accuracy_encodings}
\end{figure}
\par In Figure \ref{fig:test_accuracy_encodings} we plot the test accuracy of the trained classifiers against the number of CNOT gates in the classifier circuit. The red circles show the accuracy for classifiers with $1$--$6$ layers trained on states using amplitude encoding, the blue diamonds show the accuracy for classifiers with $1$--$7$ layers trained on FRQI states, and the green squares show the accuracy for classifiers with $1$--$6$ layers trained on NEQR states, where the different levels of color saturation correspond to the different numbers of color qubits used in the NEQR encoding. For amplitude encoding and FRQI encoding the resulting accuracy is very similar, with FRQI encoding performing slightly better at the cost of an additional qubit. Both perform significantly better than the NEQR encoding, which also requires more qubits to be implemented.
\begin{figure}[t]
\centering
\includegraphics[width=\linewidth]{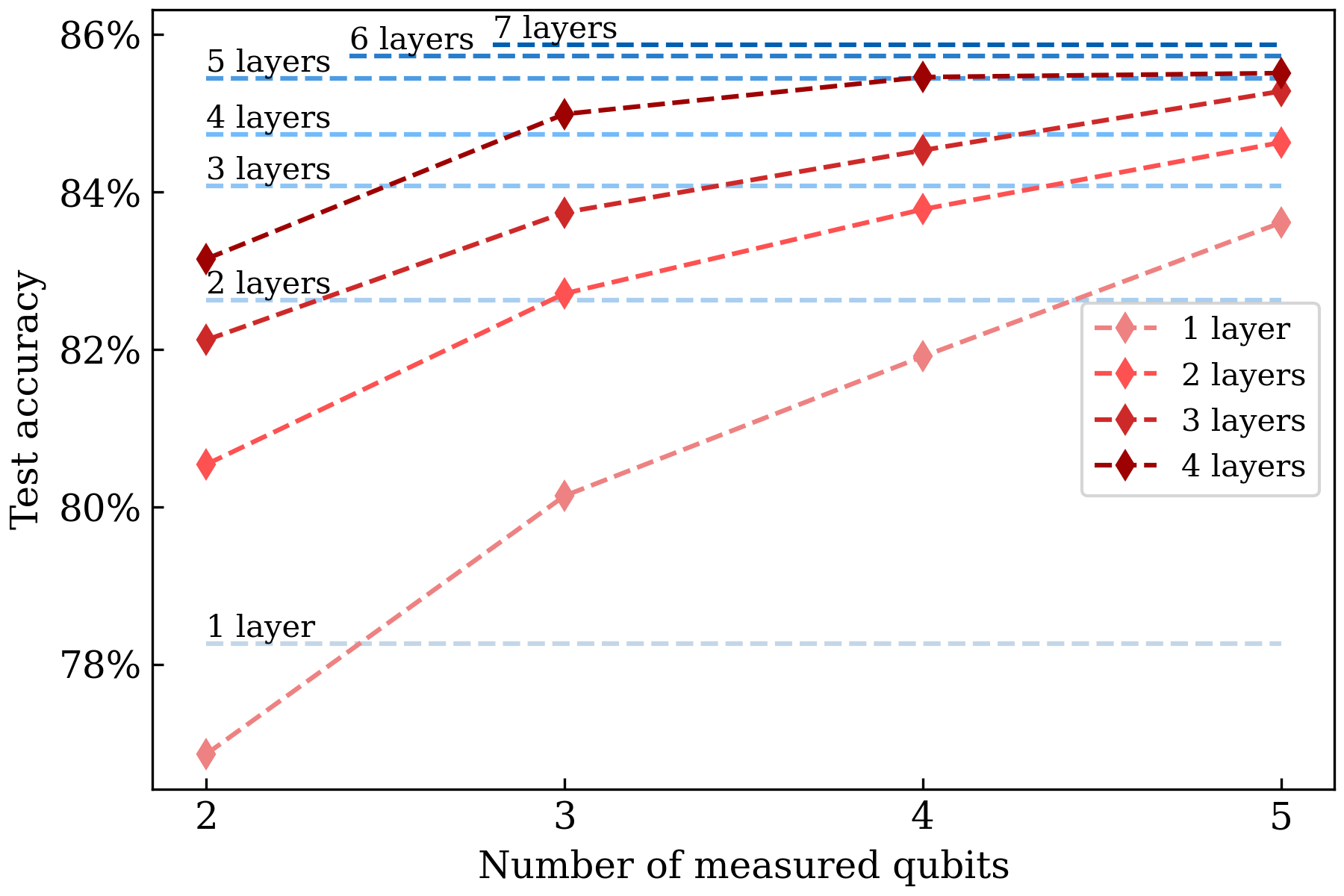}
\label{fig:test_accuracy_postprocessing}
\caption{Comparing the test accuracy of one-dimensional sequential circuits on the FRQI-encoded uncompressed Fashion-MNIST dataset with and without classical post-processing. Using classical post-processing, we can in principle choose the number of qubits $m$ we measure at the end of the circuit, as we multiply the outcome probabilities by a $k\times2^m$ matrix to assure $k$ different output values. The red data points show the test accuracy of the classifier using post-processing for different numbers of qubits measured, and the different saturations of red show the results for the different layers of the classifier circuit. As a comparison, the blue dashed lines show the results of the classifier without post-processing, where always 4 qubits are measured. Using classical post-processing, circuits with one layer fewer achieve about the same accuracy as circuits with an additional layer that do not use the post-processing.}
\end{figure}
\par Next, we consider the effect of changing the observables we measure from simple bitstrings to the more general weighted sums of bitstrings as explained in the main text. This means we are now no longer restricted to measuring 4 qubits to get at least ten different outcomes, but we can measure fewer or more qubits, as we can always get exactly ten outcomes by taking ten different weighted sums of all outcomes. For simplicity, we use the same circuit structure as before (shown in Figure~\ref{fig:1d_sequential_classifier}), and just change from measuring the last four qubits to measuring the last $m$ qubits. There are $2^m$ outcome probabilities, which we then multiply by a $k\times2^m$ matrix and add to a $k$-dimensional bias vector, giving us $k$ different observables. We assign the label to the corresponding observable that has the largest value (as in the main text). The $k\times2^m$ matrix as well as the bias vector are learned during training, together with the gates in the circuits. For the training, we initialize the upper-left $k \times k$ corner of the matrix as an identity and then add Gaussian noise to the whole matrix, and the bias is also initialized by Gaussian noise. 
\par For training the circuit we use the same procedure as described before. In Figure~\ref{fig:test_accuracy_postprocessing}, we compare the resulting test accuracy achieved on the exactly FRQI-encoded Fashion-MNIST dataset when using post-processing to the test accuracy achieved without using it. The red diamonds show the test accuracy of the trained classifier using post-processing, plotted against the number of qubits that are measured after the circuit. The different saturation levels correspond to the different number of layers of the variational circuit. For comparison, we also plot the test accuracies of the classifiers that do not use post-processing as dashed blue lines. This is the same data as was shown in Figure~\ref{fig:test_accuracy_encodings} for the FRQI-encoded images. Note that in this case the number of measured qubits is always fixed to four. We can see that the circuits using the classical post-processing need roughly one fewer layer to achieve the same classification accuracy if the number of measured qubits is four, and they can match the classification accuracy of the circuits without post-processing while measuring only three qubits instead of four. 
\par Since we now also train classical parameters to further process the results from the quantum computer, the question arises how much of the task is learned by the classical resources, and how essential the quantum part remains for the full classification task. To address this question, we remove the classification circuits and directly apply the post-processing units to the measured bitstring counts on the last $m$ qubits of the encoded image data. In that case, the classification accuracy drops to $26.98\%$, $30.22\%$, $46.53\%$ and $60.11\%$ when measuring the last $2$--$5$ qubits. We see that when we are measuring a large part of the system (i.e., $5$ of $11$ qubits), a lot of information can already be extracted classically; however, to achieve competitive results the quantum circuits remain essential.

\begin{figure}[t]
    \centering
    \begin{tabular}{l}
        (a)\\[-1.35em]
        \hspace{0.05\linewidth}\includegraphics[width=0.95\linewidth]{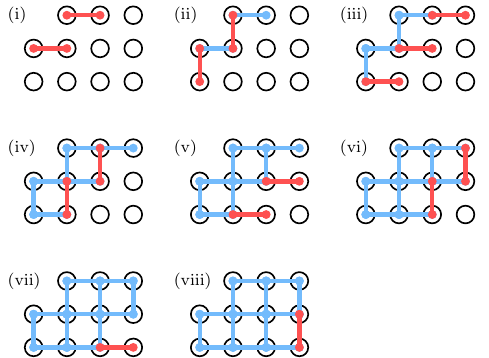}\\[2em]
        (b)\\[-1.35em]
        \hspace{0.05\linewidth}\includegraphics[width=0.95\linewidth]{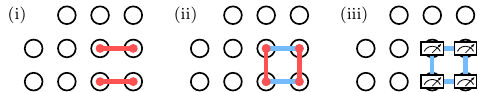}
    \end{tabular}
    \caption{A two-dimensional sequential circuit as a variational classifier. (a) A single layer of a two-dimensional sequential circuit~\cite{wei_sequential_2022, jobst_efficient_2023} for eleven qubits can be constructed in eight steps---at each step, the red gates show the newly applied gates, and the blue gates show the previously applied gates. To create deeper circuits, several layers of this sequential circuit can be applied successively, just as in one dimension. Note that as several layers are stacked, many of the gates can be applied in parallel. (b) Just as for the one-dimensional sequential circuit, we also apply a readout layer before the final measurement to the two-dimensional sequential circuit. The qubits that are finally measured are marked with the measurement symbol.}
    \label{fig:2d_sequential_classifier}
\end{figure}
\par Finally, we want to compare the effect of using different circuit ansätze for the variational classifier on the test accuracy. For this, we compare the sequential circuits we have been using so far, a two-dimensional generalization of the sequential circuits~\cite{wei_sequential_2022}, a multiscale entanglement renormalization ansatz (MERA)~\cite{Vidal2008} circuit and a quantum convolutional neural network (QCNN)~\cite{Grant2018, cong_quantum_2019}. For this comparison, we again limit ourselves to measuring ten different bitstrings and comparing their probabilities, instead of the more general setup of weighted bitstrings that we considered in the main text, and we only consider the uncompressed FRQI-encoded version of the Fashion-MNIST dataset. The one-dimensional sequential circuit is the same as in Figure~\ref{fig:1d_sequential_classifier}. For the two-dimensional generalization of the sequential circuits, we arrange the $11$ qubits in a $4\times3$ grid, leaving out the upper left corner. With this setup, Figure~\ref{fig:2d_sequential_classifier}a shows how to construct a single layer of the two-dimensional sequential circuit~\cite{jobst_efficient_2023, wei_sequential_2022}. At each step, the red gates show the newly applied gates, and the blue gates show the ones that have been applied previously. The sequentially repeating pattern here consists of gates along a diagonal, which are applied alternately horizontally or vertically. To create more expressive circuits, we can apply several layers of this ansatz successively. Note that when doing this, many gates of the circuit can be applied in parallel, such that for each layer after the first only the final four steps contribute to the circuit depth. As for the one-dimensional version of the sequential circuit, before measuring the qubits we apply a readout layer. This readout layer is shown in Figure~\ref{fig:2d_sequential_classifier}b, where also the four qubits that are measured in the end are marked by a measurement symbol.
\par
\begin{figure}[t]
    \centering
    \includegraphics{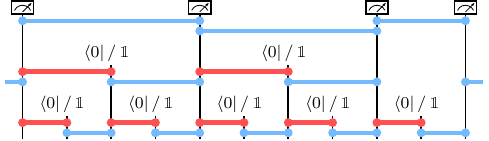}
    \caption{The circuit used for implementing a MERA and QCNN classifier. For each layer, the qubits are grouped into pairs and then first a gate is applied between neighboring pairs of qubits (the blue gates) and then within the same pair of qubits (the red gates). After each layer half of the qubits are discarded---for the MERA classifier that means measuring those qubits and post-selecting on the $\ket{0}$ outcome, such that the red gates effectively implement an isometry; for the QCNN classifier it means to not do anything to these qubits anymore, effectively tracing them out.}
    \label{fig:QCNN_classifier}
\end{figure}

 We also consider hierarchical circuits like the MERA and the QCNN~\cite{Vidal2008, Grant2018, cong_quantum_2019}. Both ans\"atze use the same circuit structure, which is shown in Figure~\ref{fig:QCNN_classifier}. In each layer, the qubits are grouped into nearest-neighbor pairs, and gates are applied first between different pairs (the blue gates) and then within a pair (the red gates). To make the classifier hierarchical, after each layer one of the qubits in each pair is discarded and the remaining qubits are treated as the new system of half the size. For the MERA circuit discarding the qubits means measuring them and post-selecting on the $\ket{0}$ outcome (denoted by $\bra{0}$ in the figure). This way, the red gates implement an isometric matrix instead of a unitary matrix, just as in the originally proposed MERA tensor network for classical computers~\cite{Vidal2008}. On a quantum computer, this post-selecting would generically be exponentially costly, so we also consider a QCNN version of the circuit, where discarding a qubit means just not applying any gates to it anymore, effectively tracing out the qubit at this step (denoted by $\mathbbm{1}$ in the figure).
\begin{figure}[b]
    \centering
    \includegraphics[width=\linewidth]{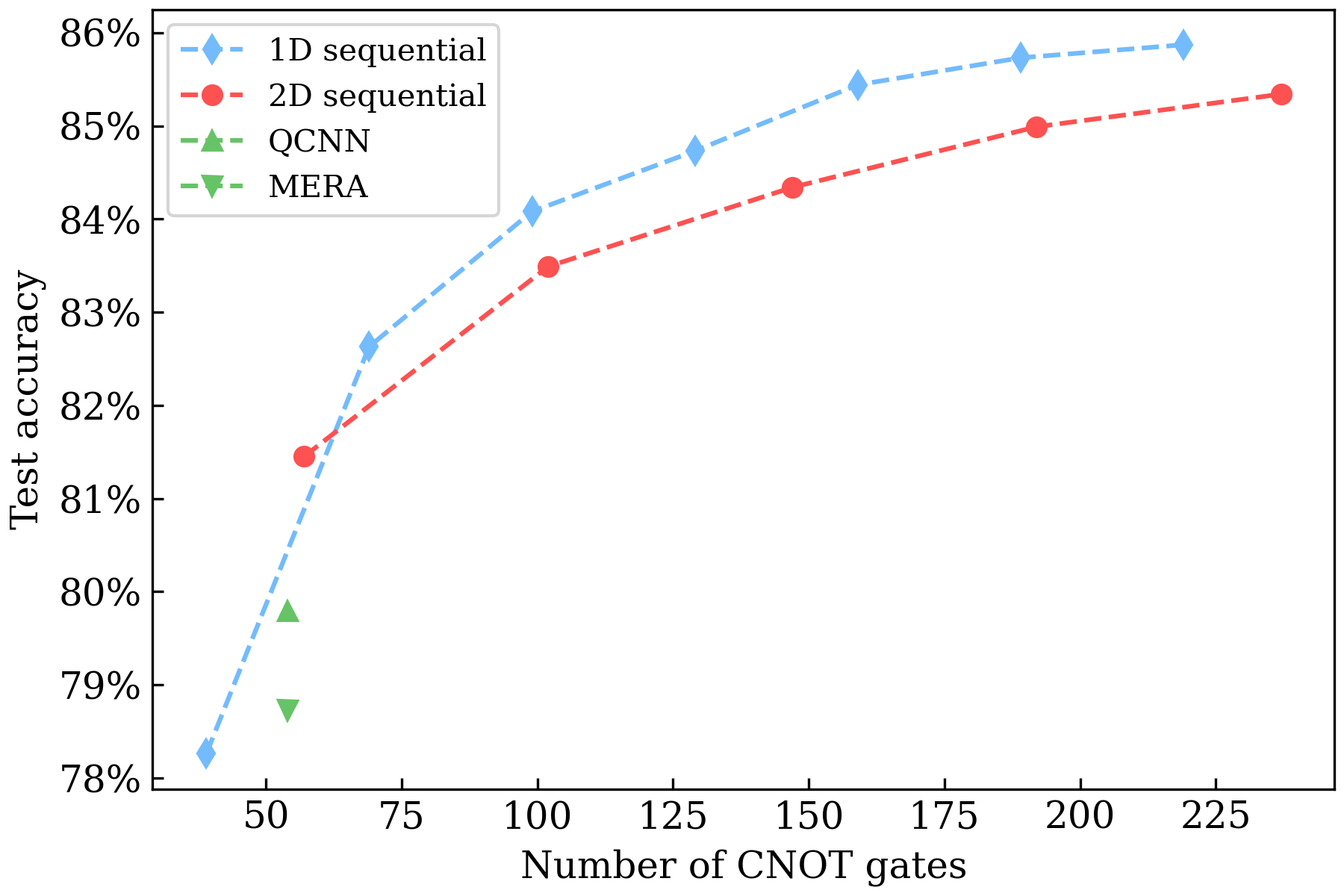}
    \caption{Comparing the test accuracy of different circuit architectures as variational classifiers on the uncompressed FRQI-encoded Fashion-MNIST dataset. We consider one- and two-dimensional sequential circuits (see Figures~\ref{fig:1d_sequential_classifier} and~\ref{fig:2d_sequential_classifier}), as well as a MERA and a QCNN circuit (see Figure~\ref{fig:QCNN_classifier}). The test set accuracy is plotted against the number of CNOT gates in the classifier circuit; note that here we count the long-range CNOT gates in the MERA and QCNN circuits as a single CNOT gate, which would be correct for quantum computers with all-to-all connectivity, but for quantum computers with nearest-neighbor connectivity such as ibmq-kolkata these would need to be implemented with a series of SWAP-gates, generically introducing a lot more CNOT gates. (A similar problem occurs for the two-dimensional sequential circuits if the hardware layout does not support a square lattice with nearest-neighbor connectivity.)}
    \label{fig:test_accuracy_classifier}
\end{figure}
\par Figure~\ref{fig:test_accuracy_classifier} shows the results for the test accuracy of the different classifiers plotted against the number of CNOT gates used in the circuit. The blue diamonds show the results for the one-dimensional sequential circuit for $1$--$7$ layers, the red circles show the results for the two-dimensional circuits for $1$--$5$ layers, the right-side-up green triangle shows the result for the QCNN circuit, and the upside-down green triangle shows the result for the MERA circuit. Note that while the one-dimensional sequential circuit with nearest-neighbor connectivity can be directly implemented on most hardware, the two-dimensional square-lattice layout with nearest neighbor-connectivity and the hierarchical circuits with long-range connectivity cannot always be directly implemented, and may need to be compiled with a series of SWAP gates. For example, both circuits could be directly implemented on all-to-all connected quantum computers, while the heavy hex layout of ibmq-kolkata would not allow for direct implementation of either. For the purpose of counting the number of CNOT gates, we assume a best-case scenario with perfect connectivity for the given ansatz, such that any two-qubit gate can be implemented using only 3 CNOTs. Even with this assumption, we see that the one-dimensional sequential circuits outperform the other circuit architectures slightly, while the overall trend is very similar as the number of resources is increased.
\end{document}